# Resistance to TB drugs in KwaZulu-Natal: causes and prospects for control

K. Wallengren,[1*] F. Scano,[2] P. Nunn,[2] B. Margot,[3] S. Buthelezi,[3] B. Williams,[4] A. Pym,[5] E. Y. Samuel,[6] F. Mirzayev,[2] W. Nkhoma,[7] L. Mvusi,[8] and Y. Pillay[8]

1. Freehouse International Ltd., Gibraltar
2. StopTB, World Health Organization, Geneva, Switzerland
3. Department of Health, KwaZulu-Natal, Pietermaritzburg, South Africa
4. South African Centre for Epidemiological Modelling and Analysis, Stellenbosch, South Africa
5. Medical Research Council, Durban, South Africa
6. Academic Complex Business Unit, Inkosi Albert Luthuli Central Hospital, National Health Laboratory Service, KwaZulu-Natal, South Africa.
7. World Health Organization, Regional Office for Africa, Harare, Zimbabwe
8. Department of Health, Pretoria, South Africa

* Correspondence to kristina@wallengren.org

## Abstract

**Background**

In 2005 there was an outbreak of XDR (extensively drug resistant) TB in Tugela Ferry, which is served by the Church of Scotland Hospital (COSH), in the uMzinyathi District, KwaZulu-Natal, South Africa. An investigation was carried out to determine if XDR TB was occurring elsewhere in the province, and to develop hypotheses for the rise in drug resistance with a view to developing a strategy for the control of MDR (multi-drug resistant) and XDR TB in the province and elsewhere.

**Methods**

TB incidence and treatment success rates, for each of the 11 districts in the province, were obtained from the provincial electronic TB register for the years 2002–2007. The results of culture and drug sensitivity tests for the years 2002 to 2007 in each of the districts were compiled and culture taking practices were compared to the number of MDR TB cases. Interviews were conducted with key personnel in affected sites.

**Findings**

In 2007, 2799, or 2.3% of 119,218 notified TB cases in the province were multi-drug resistant (MDR), and of these 270 (9.6%) were XDR. The two worst affected districts were uMzinyathi where 226 (4.1%) of 5522 notified TB cases were MDR, and of these 120 (53%) were extensively drug resistant (XDR), and uMkhanyakude where 337 (4.8%) of 6991 notified TB cases were MDR, but of these only four or (1.2%) were XDR. The worst affected medical centre was COSH where 164 or 9.8% of notified TB cases were MDR and of these 99 (60%) were XDR.

**Interpretation**

Very high rates of XDR TB in the province are only found in uMzinyathi district even though MDR TB is common in most other districts. XDR may arisen at COSH because of the early and effective integration of the TB and HIV programmes in overcrowded and poorly ventilated facilities particular to COS.H To control XDR TB better management of both susceptible and resistant forms of TB is needed including treatment supervision, infection control and HIV management.

## Introduction

Over the last ten to twenty years resistance to TB drugs has increased and in some countries up to 20% of new TB cases are infected with multi-drug resistant (MDR) strains of TB that are resistant to at least isoniazid (INH) and rifampicin (RIF).[1] In 2005 an outbreak of extensively drug resistant (XDR) TB was reported from Church of Scotland Hospital (COSH), Tugela Ferry, in the KwaZulu-Natal Province of South Africa.[2] These XDR patients were resistant to isoniazid and rifampicin but also to ethambutol (ETA), streptomycin (STP), aminoglycosides (kanamycin) and fluoroquinolones (ofloxacin). The initial investigation found that of 475 patients with confirmed culture positive TB 185 (39%: 35%–44%; here and elsewhere errors are 95% confidence limits) were MDR and that of these 30 (16%; 11%–22%) were XDR. Forty four XDR TB patients were tested for HIV and were all HIV-positive. Their median survival time, after diagnosis, was 16 days.[2] The unexpected and rapid emergence of XDR TB, the association with HIV and the very high mortality was of great concern; if similar outbreaks were to occur more widely in the province, the country or the region they could lead to considerable mortality, jeopardize TB control programmes, and be very costly to contain.

In 2007 the KwaZulu-Natal Department of Health initiated an investigation of MDR and XDR TB in the province to establish a) the geographical extent of the epidemic; b) the reasons why it happened where and when it did; c) the nature of the association with HIV; and d) ways to contain and control the epidemic.

We collected, collated and analyzed data from districts, hospitals and laboratories on: a) laboratory testing; b) culture taking practices; c) the prevalence of HIV; d) TB programme performance; and e) the management of MDR. We focussed on the years from which TB culture data was available, 2001 to 2007, covering the period from four years before, until two years after, the start of the outbreak.

## Methods

### TB culture

Between 2001 and 2007 TB cultures and drug sensitivity testing (DST) in KwaZulu-Natal were done in laboratories at the King George V (KGV) and Inkosi Albert Luthuli Central (IALCH) Hospitals. In 2006 the laboratory at KGV was closed and from March 2006 IALCH took responsibility for all cultures and DST testing in the province. TB culture and sensitivity data were available from KGV laboratory from 2001 to 2005, and at the IALCH laboratory from 2002 to 2007 including: hospital number, patient name, sex, specimen number, type of sample (sputum, gastric washing and spinal fluid), referring centre, date received, and drug sensitivity. The hospital number, name, age and sex were used to clean the data and remove duplicates. When the same case had more than one culture, the result was linked to the first culture test and classified according to the highest level of drug resistance if the results differed.

### Drug resistance testing

Prior to 2000, all samples were tested for sensitivity to isoniazid, rifampicin and ethambutol. If resistance was detected to any one drug, further testing was done for resistance against streptomycin, kanamycin and ciprofloxacin. In July 2004, the policy at IALCH was expanded to include routine testing of all samples for all six of the above drugs since rates of resistance were increasing and this approach reduced the turnaround time for second line DST. The proportion of samples tested for sensitivity to all six drugs increased from 27% to 49%, 63% and 100% in successive years from 2003 to 2006. Ciprofloxacin testing was replaced by ofloxacin in 2004 but the database at IALCH was not changed, and from 2004 onwards 'ciprofloxacin resistance' is actually 'ofloxacin resistance.' TB cultures were prepared using the MGIT 960 liquid culture system followed by drug susceptibility testing (DST) on solid Middlebrook 7H10 mini-plates.[3,4]

Drug resistance testing at KGV was performed at the referring doctor's request. In 2004 and 2005 KGV did not test for resistance to fluoroquinolones.

### Culture taking

We compiled the data for liquid MGIT samples that were cultured in each district from 1 April 2006 to 31 March 2007. According to the national guidelines, cultures should be taken at the initial investigation of all retreatment cases, treatment failures, sputum smear-negative cases with clinical signs of TB who do not respond to antibiotics, and all HIV-positive TB suspects.[5] We used the Electronic TB Register (ETR), maintained by the National Department of Health, to estimate the number of TB cultures that should have been done in each district and compared this with the number that were done.

### HIV

The prevalence of HIV in each district was taken from annual antenatal clinic sero-prevalence surveys, 2003 to 2006.[6-9] While these data may tend to overestimate HIV prevalence in the general population, they provide a reasonable means of comparing the HIV burden in each district.

### TB notification data

The number of new and retreatment TB cases notified each year from 2002 to 2007 and the treatment outcomes by district and health facility from 2002 to 2005 were taken from the ETR.

### Population data

The population of each district and each hospital catchment area was provided by the Geographical Information Systems (GIS) Component of the Provincial Department of Health.

### MDR TB treatment

KGV was set up as a TB hospital in the 1950s with over one thousand beds. With the passage of time and deterioration of the buildings the number of beds was reduced and in 2007 only 160 beds were available for treatment of TB patients. KGV served as a specialised referral hospital for MDR TB patients and, at the time of the study, was the only place in KwaZulu-Natal where MDR TB treatment was available. For the period 1998 to 2007 available patient data included hospital number, name, age, sex, referring centre, drug sensitivity, date admitted and discharged, treatment outcomes, and HIV status. MDR outcomes were not available for the 2006 and 2007 cohorts, as patients had not completed treatment at the time of the investigation. We compared the number of patients in the province that tested positive for MDR with the number that were admitted to KGV between 1 July 2006 and 30 June 2007 to assess the proportion of MDR patients that were admitted for treatment.

### Interviews with health service personnel

Interviews were conducted with clinical and administrative hospital staff working in the 1990s and 2000s at major hospitals in the two districts where MDR burden was the highest COSH in uMzinyathi, and Manguzi and Hlabisa hospitals in uMkhanyakude where XDR was the lowest, as well as a reference hospital, Edendale, in uMgungundlovu district. Twelve interviews, each lasting for up to one hour, were carried out with the Medical Manger, the Chief Medical Officer, and the Chief Professional Nurse (TB coordinator, DOTS nurses, or infection control nurses).

**Provincial map**

A map of the province with the districts indicated for reference is given in the Appendix.

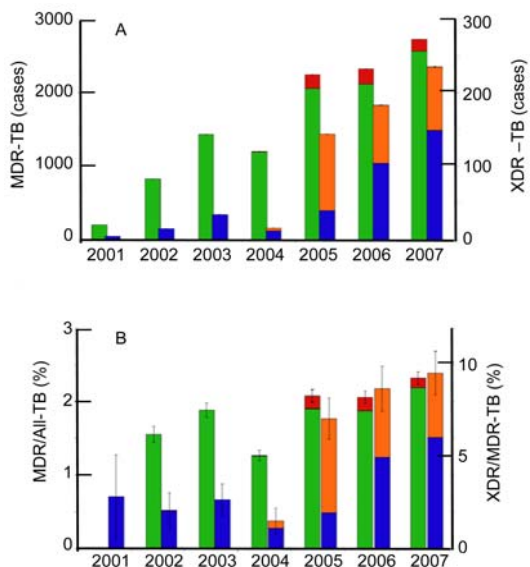

Figure 1. The number (A) and proportion (B) of MDR and XDR cases in KwaZulu-Natal. Green and red: MDR, red MDR from COSH; blue and orange XDR, orange XDR from COSH.

## Results

**Time trends**

The TB notification rate in KwaZulu-Natal rose from about 54,000 in 2002 to over 119,000 in 2007. The number of MDR TB cases rose more quickly from 210 in 2001 to 2799 in 2007 (Figure 1A, green and red bars) by when 2.3% of all TB cases were MDR (Figure 1B, green and red bars). While some XDR cases were identified before 2005, the numbers rose very rapidly in 2005, and more slowly after that, with about half of all XDR cases being reported from COSH (Figure 1A blue and orange bars) in 2007. The proportion of MDR cases that were XDR increased from 2.1% in 2002 to 9.6% in 2007 (Figure 1A green and red bars).

The substantial increase in the proportion of XDR cases followed the discovery of XDR TB at COSH in February 2005 after which case finding was intensified and cultures were done on all TB suspects at COSH. In 2006 other health facilities in the province began to request cultures and the proportion of MDR cases that were XDR, from outside COSH, doubled between 2005 and 2006 (Figure 1B, blue bars). Much of the increase in the number of MDR and XDR cases is the result of the increase in the number of cultures that were done. In 2007, 57% more samples were cultured in the province compared to 2006. Unfortunately, the total number of samples cultured in the province was not available for 2005 or earlier.

**Variation among districts**

Figure 2 shows the overall TB notification rate, the proportion of cases that were MDR (including XDR) and the proportion of these that were XDR, averaged over the years 2006 and 2007. The notification rate for all forms of TB ranged from 750 to 1400 per 100 000 population per year (blue bars) whereas the proportion of cases that were MDR ranged from 1% to 5% (red bars). However, 46% of MDR cases were XDR in uMzinyathi, more than three times the proportion in uMgungundlovu, the next highest district, where 14% of MDR cases were XDR.

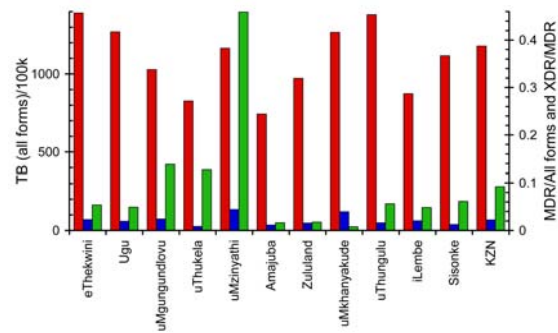

Figure 2. TB notification rates: all forms (red), MDR as a proportion of all forms (blue) and XDR as proportion of MDR (green) for the districts in KwaZulu-Natal for 2006-2007.

*Culture taking*

Culture taking practice varied greatly among districts. The cities of Durban in eThekwini and Pietermaritzburg in uMgungundlovu have specialized TB hospitals. The number of cultures done in these districts was 1.4 and 1.0 per TB patient, respectively. In the remaining districts the culture rate was much lower ranging from 0.08 per TB patient in Amajuba to 0.58 per TB patient in uMzinyathi. Excluding eThekwini and uMgungundlovu the high burden of MDR TB in uMzinyathi and uMkhanyakude is not significantly correlated with the proportion of TB patients whose sputa were cultured ($\rho = 0.451$; $p = 0.19$).

*HIV prevalence*

In 2006 39.0% ± 1.2% of women attending antenatal clinics in KZN were HIV-positive and there was no significant variation among districts apart from uMzinyathi where the prevalence, 27.9% (23.0%−32.8%), was significantly below the provincial average.

**Treatment outcomes**

*All TB cases*

In 2005 the treatment success rate for TB in the province was 63% among new and 50% among retreatment cases, well below the 80% target. The treatment success rate was highest in uMzinyathi at 71% in new and 54% in retreatment cases. Default rates were high in eThekwini (23%) and uThukela (18%), death rates were high in Amajuba (15%), Zululand (14%), uMzinyathi (13%), and uMkhanyakude (10%), and the proportion 'not evaluated' was high in uMgungundlovu (34%), Ugu (28%) and uThungulu (21%). The reported treatment success rates at COSH were low, at only 27%, but 68% of their patients were transferred out to primary health care facilities within the catchment area of the hospital and reported as such. Since the first-line treatment outcomes (71% treatment success) for the district (uMzinyathi) are among the best in the province, poor treatment outcomes are unlikely to explain the outbreak of XDR in the province.

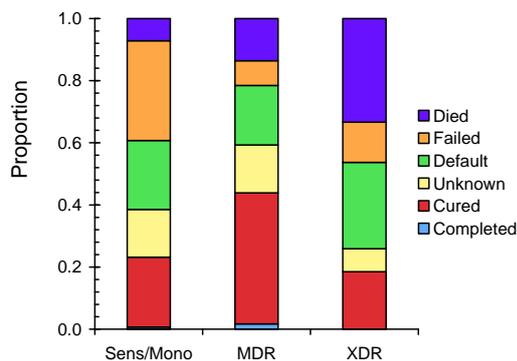

Figure 3. Treatment outcomes for 729 patients admitted to KGV with fully sensitive or mono-resistant TB, 2315 with MDR TB and 54 with XDR TB.

*MDR and XDR treatment outcomes*

Figure 3 gives the treatment outcomes for 3098 TB patients admitted to KGV between 1994 and 2005. XDR patients were more likely than MDR patients to die (OR 1.45; 0.76−2.79, not significant) and less likely to be cured (OR 0.44; 95% CL 0.21−0.94).

However, the treatment success rate for MDR patients from uMzinyathi was 12.0% (5.5%−18.5%), considerably below the provincial average of 44%, while in uMkhanyakude it was 38.4% (34.5−42.2%), only slightly below the provincial average. The low treatment success rates in uMzinyathi were associated with higher than average default rates, 28.0% (19.0%−37.0%), and death rates, 32.0% (22.7%−41.3%) but lower than average failure rates 12.0% (5.5%−18.5%). We therefore examined the treatment outcomes more carefully in the four hospitals in uMzinyathi.

*MDR and XDR in uMzinyathi*

The number of MDR TB cases as a proportion of the catchment population varies among the four hospitals in uMzinyathi District and, as shown in Figure 4, is significantly higher in COSH than in the other three: Charles Johnson Memorial (CJM), Dundee (DUN), or Greytown (GTN) Hospitals. However, in 2006, COSH took 1,300 cultures per catchment population, 26 times as many as in Dundee where they took 50 cultures per catchment population and this could explain the different rates within the district.

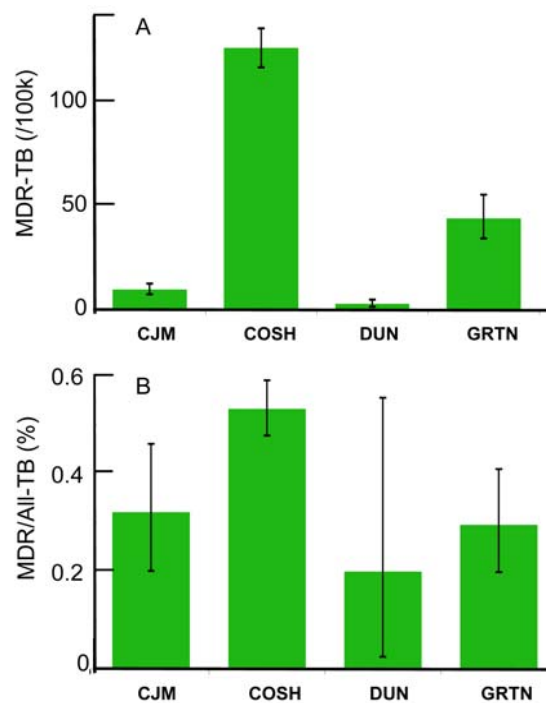

Figure 4. A: The number of MDR cases per 100k catchment population (left) and B: the proportion of MDR cases that were XDR (right) for patients tested at IALCH from 2005 to 2007. CJM: Charles Johnson Memorial; COSH: Church of Scotland; DUN: Dundee; GRTN: Greytown (GTN) Hospitals.

There was less variation in XDR as a proportion of MDR than in MDR as a proportion of all TB patients among the district hospitals as this depends less on the proportion of patients whose sputa are cultured. The proportion of MDR cases that were XDR in COSH is 2.4 times the average of the other three in 2005-2007.

*MDR TB treatment*

Between 1998 and 2007 the number of MDR-TB cases admitted to KGV increased by a factor of eight, from 147 to 1086, while the number of XDR TB cases increased from less than two cases per year up to 2004, to 31 in 2005 and 148 in 2007. Because the number of beds did not increase the average length of stay fell from six months in 1998 to two months in 2007, excluding those who died in the hospital. This led to an increase in the proportion of MDR TB patients with a positive smear or culture on discharge; between 2000 and 2005, the proportion of patients who were smear positive on discharge fell from 15% to 5% but rose again to 17% in 2006 while the proportion who were culture positive on discharge rose from 25% in 2005 to 33% in 2006.

The prevalence of HIV among patients admitted to KGV was high in both MDR (72.4%; 69.5%–75.4%) and XDR patients (76.2%; 68.8%–83.6%) but a recent estimate suggests that 73% of all TB cases in South Africa are HIV-positive.[1] These numbers do not suggest a strong association between HIV and MDR or XDR TB.

*MDR TB cases treated at KGV*
In 2006, only 30% of laboratory identified MDR cases in the province were admitted for MDR TB treatment and in order to cope with the increasing patient load, another 16% of all MDR cases were initiated on MDR TB treatment as outpatients with little or no support in their home areas which could be hundreds of kilometres away and with patients having to return monthly to KGV to collect medication. The remaining 54% (1505 MDR TB cases) had either died before receiving their diagnosis or remained untreated. In the same year the average time between sputum collection and admission to KGV was 16 weeks; on average it took three to six weeks to culture specimens and do DST, three weeks to return results to the referring health facility, three weeks to trace patients, and four to eight weeks to admit patients to KGV.

In 2005 the treatment success for all MDR patients at KGV was 40% but, given that over 70% of patients were infected with HIV, it is no worse than in other similar settings. However, uMzinyathi district had the highest default rates from MDR TB treatment in the province between 1996 and 2004.

*Potential for the spread of MDR TB from KGV hospital*
The highest MDR burdens as proportion of TB cases are in uMkhanyakude (4.8%) and uMzinyathi (4.1%) but the former has the lowest XDR rates in the province (1.0%) while the latter has the highest (46.8%). In the ten-year period from 1994 to 2004, prior to the outbreak of XDR TB in Tugela Ferry, 167 patients from uMkhanyakude compared to 26 patients from uMzinyathi were admitted to KGV for MDR treatment. The risk for developing, or contracting and spreading XDR was therefore higher in patients from uMkhanyakude than uMzinyathi. While it seems unlikely that poor treatment at KGV was the main reason for the outbreak seen in Tugela Ferry the rising demand for treatment at KGV without the necessary resources has increased the proportion of MDR TB patients being discharged while still infectious in 2006 when 15% were smear positive and 46% were culture positive.

## Discussion

**Drug resistance burden**

The first two registered XDR TB patients in South Africa were admitted to KGV hospital for treatment in 2000 and both came from Durban (eThekwini). In 2007, in KZN 2.3% of all TB cases were MDR, close to the estimates from the 2002 National Drug Resistance Survey which gave 1.8% in new and 5.5% in retreatment cases.[10] While the drug sensitivity data from KZN after 2006 is fairly reliable, under-sampling may have led to under-estimates of the true number of drug-resistant patients. There may also be less variation among districts than suggested by the available data because districts that reported a low MDR TB burden also took fewer samples for culture. Among the district hospitals within uMzinyathi, COSH had the highest MDR TB burden but also referred the greatest proportion of cases for culture. But, even accounting for differences in culture taking practices, MDR TB rates are particularly high in uMzinyathi, where XDR was first discovered in Tugela Ferry, and in uMkhanyakude, the coastal district that borders Mozambique.

XDR TB is present in all districts of the province but is more common in uMzinyathi than in the other districts in KZN. Within uMzinyathi, the extreme XDR burden is not unique to COSH and has probably spread to the other district hospitals, although COSH remains the worst affected. It seems that XDR TB has not spread much to other districts and in particular uMkhanyakude district has the lowest XDR TB burden (1.2% in 2007) in the province but the highest MDR TB burden (4.8% in 2007).

After 2004 XDR TB as proportion of MDR TB provides a robust indicator of trends in XDR TB for two reasons: 1) routine DST for first and second line TB drugs has been performed on the majority of samples since 2004, and 2) the XDR/MDR ratio is less dependant on the culture taking practice.

**Culture taking practice**

Using the expected number of cultures entails several assumptions regarding implementation of guidelines and estimates of cases requiring cultures. Using the population or reported TB burden as the denominator is also imprecise due to changes in case finding and reporting. However, the analysis presented here using any one of the three denominators gave similar results and gives some confidence in the results of the comparison.

The strength of the analysis based on culture taking practice is that the data used in the numerator was cleaned and represents cases rather than samples. While the total number of samples cultured increased over the years, particularly after 2005, it was not possible to determine the time trends per district for culture taking practice.

**Possible reasons for the emergence of drug resistance**

*Suboptimal TB treatment*
By 1996, the South African National TB Control Programme (NTCP) was functional and based on the DOTS strategy. However, KZN did not adhere to the NTCP guidelines and used individualised treatment based on drug-sensitivity testing and DOTS was not promoted. In 2001, single drugs were replaced by fixed-dose combination drugs to ensure standardized treatment, and individualised treatment based on DST results was abandoned, which may have contributed to development of drug resistance in the province[11] and treatment success rates in 2003 were low.

Particular circumstances may have allowed MDR development to reach high levels in uMkhanyakude and uMzinyathi. In several hospitals in uMkhanyakude, TB treatment was given two to three times per week with four drugs for six months without a separate intensive and continuation phase until 2001 when single drugs were replaced by fixed dose combination drugs. In uMzinyathi, because the geographical conditions made access to the hospital difficult, COSH regularly prescribed six-months TB treatment to spare patients from having to return monthly to collect further treatment although patients were followed up after two and six months.

The development of XDR TB implies suboptimal use of second line drugs. In uMzinyathi, a two-week course of ciprofloxacin was used only at COSH in the period 1998-2004 to treat HIV-positive patients suffering from diarrhoea. In the rest of the province, ciprofloxacin was used in a one-day course of 500 mg for the syndromic management of gonorrhoea. Any other use of ciprofloxacin is restricted due to provincial regulations. Kanamycin and amikacin were rarely used in the province except for the treatment of TB.

*HIV prevalence and poor TB programme*
High HIV prevalence and poor TB programme performance could also have contributed to development of drug resistance in a particular area. However, HIV prevalence was no higher in uMzinyathi than in any other district; similarly, TB programme performance in uMzinyathi between 2002 and 2005 was not worse than in other districts. Hence, neither HIV prevalence nor poor TB programme performance are likely to explain the exceptionally high rates of drug resistant TB in uMzinyathi although both may contribute to the overall problem of drug resistance in the province.

*MDR TB treatment*
The number of patients admitted to KGV has increased beyond the capacity of the hospital to treat them adequately resulting in long waiting times for treatment and a large proportion of MDR TB patients never receiving treatment. While this could have contributed to the transmission of drug-resistant TB in the province it is unlikely to have caused the surge of XDR TB in uMzinyathi. On the other hand, the high rate of defaults and transfers-out in uMzinyathi in the decade prior to the XDR outbreak may have contributed to the spread of second line drug resistance in that district.

*Lack of infection control in congregate settings*
Until very recently awareness of TB infection control has been low throughout the province. Prior to the start of the national rollout of ART in 2004 COSH ran an effective and comprehensive HIV programme (Dr. T. Moll, pers. comm.) in which HIV patients were frequently hospitalised and treated for opportunistic infections, and where TB and HIV services were integrated at facility level. However, due to the lack of adequate facilities, TB and HIV patients were brought together in waiting rooms, at VCT sites, in TB and HIV clinics, in patient education classes, and in hospital wards, all of which were overcrowded and unventilated thus facilitating the transmission of TB, including MDR and XDR TB. In other sites at the time there was much less HIV testing and ART provision, and the TB and HIV programmes were not so integrated. Furthermore, a large proportion of XDR TB appears to have been acquired through primary infection, some of it due to a significant amount of nosocomial transmission at COSH.[2]

**Conclusion**

It has been suggested that multi drug-resistant (MDR) and extensively drug-resistant (XDR) tuberculosis (TB) epidemics are rapidly expanding in South Africa and the high death rates in those with XDR generated considerable alarm.[12] However, this study reveals that, within KZN province, high rates of XDR-TB were restricted to uMzinyathi district, and COSH in particular. It may never be possible to know precisely how XDR TB arose in COSH, but a number of factors are likely to have contributed to its emergence, including the integration of the TB and HIV programmes without sufficient infection control measures, and empirical use of fluoroquinolones for relatively long periods in patients with advanced HIV infection. Other factors that are present throughout the province and which may have contributed to the emergence of XDR TB at COSH include high HIV prevalence and suboptimal TB management.

Outside uMzinyathi, the levels of MDR found in this study were sufficiently high to give rise to the concern that, together with suboptimal TB management, high levels of HIV in patients attending hospital, and insufficient infection control measures in place, XDR-TB could develop, and spread, in many other districts in the province.

A number of measures are vital to minimise the likelihood of this happening. First, surveillance to monitor MDR and XDR trends, is essential, and the policy of routine drug susceptibility testing for all culture positive isolates at IALCH is a step in the right direction, but needs to be backed up by clinicians requesting culture examination whenever clinically indicated. Second, case finding and case management of MDR needs considerable strengthening, both to ensure good support for ambulatory care of TB patients, especially for those who cannot be admitted to King George V, and also for those who are managed at KGV. Third, TB infection control must be strengthened in all health care facilities as a matter of urgency. If these improvements are made there is every reason to believe that drug-resistant TB can be managed effectively in the province.

**Acknowledgements**

The situational analysis of drug resistance in KwaZulu-Natal was prepared at the request of the Provincial Department of Health, KwaZulu-Natal, with financial and technical support of the World Health Organization and would not have been possible without the collaboration and involvement of National Health Laboratory Services, Inkosi Albert Luthuli Central Hospital TB laboratory, King George V Hospital, Church of Scotland Hospital and Department of Health. We thank Adrian Koopman for advice on the spelling of Zulu district names.

We especially thank K. Catterick, M. Chetty, C. Connolly, O. Diaz, R. Draper, K. Grimwade, E. Immelman, M. Loveday, I. Master, D. Miller, T. Moll, L. Mpontshane, Z. Ndlela, P. Nijs, G. Osburn, S. Oza, C. Pfaff, V. Raman, A. Reid, L. Roux, N. Sattar, J. Steengaard, L. Thompson, and D. Wilson for their help, work and support which made this study possible.

# Appendix

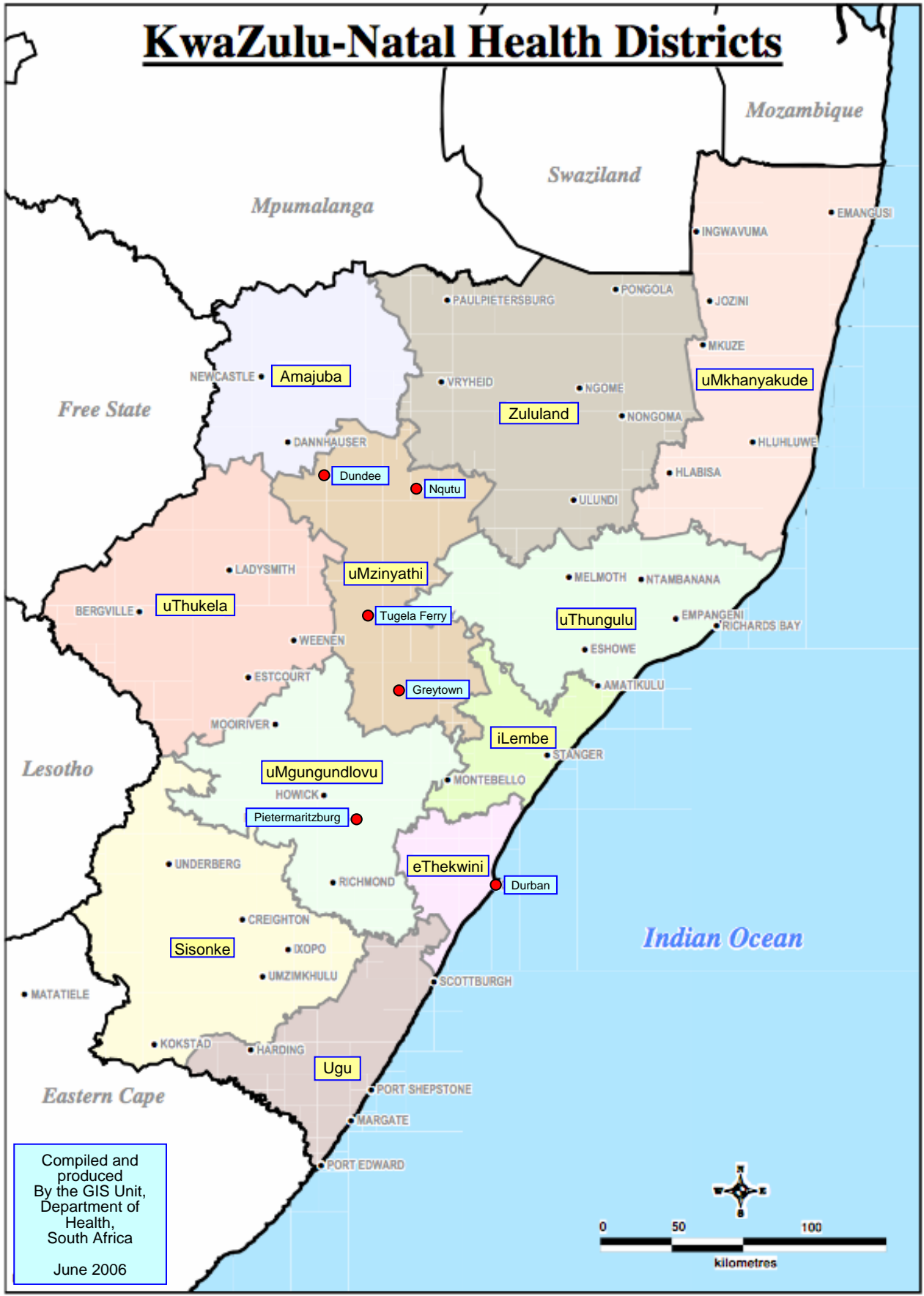